\documentclass[%
preprint,
superscriptaddress,
amsmath,amssymb,
aps,
]{revtex4-1}

\usepackage{graphicx}
\usepackage{dcolumn}
\usepackage{bm}

\usepackage{multirow}
\usepackage[table,xcdraw]{xcolor}
\usepackage{lipsum}
\usepackage{textcomp}
\usepackage{amssymb}

\begin{document}


\title{Assessment of learning tomography using Mie theory}


\author{JooWon Lim}
\affiliation{Optics Laboratory, \' Ecole Polytechnique F\' ed\' erale de Lausanne (EPFL), 1015 Lausanne, Switzerland.}
\author{Alexandre Goy}
\affiliation{Optics Laboratory, \' Ecole Polytechnique F\' ed\' erale de Lausanne (EPFL), 1015 Lausanne, Switzerland.}
\author{Morteza Hasani Shoreh}
\affiliation{Optics Laboratory, \' Ecole Polytechnique F\' ed\' erale de Lausanne (EPFL), 1015 Lausanne, Switzerland.}
\author{Michael Unser}
\affiliation{Biomedical Imaging Group, \' Ecole Polytechnique F\' ed\' erale de Lausanne (EPFL), 1015 Lausanne, Switzerland.}
\author{Demetri Psaltis}
\email{demetri.psaltis@epfl.ch}
\affiliation{Optics Laboratory, \' Ecole Polytechnique F\' ed\' erale de Lausanne (EPFL), 1015 Lausanne, Switzerland.}

\begin{abstract}
In Optical diffraction tomography, the multiply scattered field is a nonlinear function of the refractive index of the object. The Rytov method is a linear approximation of the forward model, and is commonly used to reconstruct images. Recently, we introduced a reconstruction method based on the Beam Propagation Method (BPM) that takes the nonlinearity into account. We refer to this method as Learning Tomography (LT). In this paper, we carry out simulations in order to assess the performance of LT over the linear iterative method. Each algorithm has been rigorously assessed for spherical objects, with synthetic data generated using the Mie theory. By varying the RI contrast and the size of the objects, we show that the LT reconstruction is more accurate and robust than the reconstruction based on the linear model. In addition, we show that LT is able to correct distortion that is evident in Rytov approximation due to limitations in phase unwrapping. More importantly, the capacity of LT in handling multiple scattering problem are demonstrated by simulations of multiple cylinders using the Mie theory and confirmed by experimental results of two spheres.

\end{abstract}

\maketitle


\section{Introduction}
Quantitative phase imaging (QPI) is a tool for measuring the sample-induced phase delay, which relates to the refractive index (RI) contrast. Each material has its own distinct RI value and therefore QPI can provide physiological information \cite{lee2013quantitative} such as the structure and dynamics of cells \cite{popescu2006diffraction, park2011optical}, quantification of specific molecules \cite{park2009spectroscopic, jung2014biomedical} and the dry mass \cite{barer1952interference, popescu2008optical}, which serves as an important factor for monitoring cell growth. Among other QPI techniques, optical diffraction tomography (ODT) enables us to visualize 3-D RI distributions from multiple 2-D scattered fields acquired at various illumination angles. It provides the physiological information through measuring the 3-D RI distribution without any exogenous labeling agents, making it a powerful tool for various physiological studies. 

Despite the many potential applications of ODT, there are obstacles that need to be resolved. One obstacle comes from the fact that views far from the optical axis are usually not accessible due to the limited numerical aperture of the optics. This is referred to as the missing cone problem. It causes the underestimation of RI values and the elongation of RI tomograms along the optical axis \cite{sung2011deterministic, sung2009optical}. The missing cone problem can be mediated using sparsity-based regularization algorithms \cite{sung2011deterministic, lim2015comparative}. A more fundamental problem is the inherent nonlinearity in scattering through inhomogeneous media \cite{born1999principles}. These nonlinear effects are assumed to be negligible in conventional linear ODT frameworks (Born and Rytov approximations). The validity of the approximations were well studied in showing the restriction of the linear assumption \cite{slaney1984limitations, chen1998validity}, in simple objects such as a single cell. In addition, distortions caused by multiple scattering can be severe for thick or high contrast samples especially when several scattering objects are aggregated \cite{azimi1983distortion}.

Recently, reconstruction algorithms that consider the nonlinear process have been proposed \cite{tian20153d, kamilov2015learning}. The Beam propagation method (BPM) can be used as the forward model combined with the sparsity based regularization in the iterative reconstruction scheme \cite{kamilov2015learning}. BPM is a method that describes the propagation of light through a medium. It consists of two sub-steps: Diffraction followed by a refraction step \cite{okamoto2010fundamentals}. The BPM can implement multiple scattering at different depths within the medium. Therefore, iterative reconstruction algorithms that combine the BPM and a sparsity-based regularization can outperform the conventional linear model (Rytov approximation) \cite{kamilov2015learning,kamilov2016optical}. 

However, to fairly investigate the capacity of the nonlinear model over that of the linear model, the same reconstruction framework should be used except for the forward model part. In order to distinguish the performance of LT from the effects of a sparsity-based regularization, it should be compared with the linear tomography which utilizes the same reconstruction algorithm scheme including the sparsity-based regularization. In addition, in order to quantitatively compare any improvements made by the nonlinear algorithm, we need the ground truth of the 3-D RI distribution of the object.

In this paper, the same algorithmic scheme \cite{kamilov2015learning, kamilov2016optical, beck2009fast} was used for the linear and nonlinear algorithms except for the forward model part. Simulated measurements were generated using the Mie theory whose analytical solution serves as the ground truth \cite{bohren2008absorption}. The two models are tested on various schemes where the RI contrast and the size of the sample are controlled under three different schemes: 1) variable RI contrast with fixed size, 2) variable size with fixed RI contrast, and 3) different RI contrast and size with fixed sample-induced phase delay. In addition, in order to compare the ability of each model in dealing with multiple scattering caused by multiple objects, the two models are tested using the Mie theory for multiple cylinders \cite{schafer2012calculation}. After that, the capacity of each model when phase unwrapping fails is discussed. Finally, we applied algorithms on an experimental data confirming the simulation results.

\section{Theory}
\subsection{\label{sec:level2}Optical diffraction tomography (ODT)}
The fundamental Helmholtz equation describing the scattering in inhomogeneous medium can be written as
\begin{equation}\label{eq:Helmholtz}
\nabla^2U_s(\textbf{r}) + k^2U_s(\textbf{r}) = -4 \pi F(\textbf{r})U(\textbf{r}),
\end{equation}
where $U(\textbf{r})$ is the total electric field: the sum of the incident field $U_i(\textbf{r})$ and the scattered field $U_s(\textbf{r})$. $F(\textbf{r}) = \frac{k^2}{4\pi}\left(\frac{n(\textbf{r})^2}{n_0^2}-1\right)$ is the scattering potential in a sample immersed in a medium with refractive index $n_0$. The optical wavelength in free space is $\lambda$  resulting in wavenumber, $ k = \frac{2\pi n_0}{\lambda} $ . The integral solution of Eq. (\ref{eq:Helmholtz}) can be obtained using the homogeneous Green's function   resulting in 
\begin{equation}\label{eq:solution}
U_s(\textbf{r}) = \int_{V} F(\textbf{r$^{\prime}$})U(\textbf{r$^{\prime}$})G(\textbf{r}-\textbf{r$^{\prime}$}) d\textbf{r$^{\prime}$},
\end{equation}
where $G(\textbf{r}-\textbf{r$^{\prime}$}) = \frac{e^{ik|\textbf{r}-\textbf{r$^{\prime}$}|}}{|\textbf{r}-\textbf{r$^{\prime}$}|}$ is the Green's function of 3-D Helmholtz equation, Eq. \ref{eq:Helmholtz}.

\subsection{Linear forward model}
The scattered field $U_s(\textbf{r})$ in Eq. (\ref{eq:solution}) is linear in $U_i(\textbf{r})$ but nonlinear in $F(\textbf{r})$. Born and Rytov are the two linearizations. Both of them are based on the Wolf transform \cite{born1999principles} as, 
\begin{equation}\label{eq:Wolf}
U_s^{Approx}(\textbf{r}) = \int_{V} F(\textbf{r$^{\prime}$})U_i(\textbf{r$^{\prime}$})G(\textbf{r-r$^{\prime}$}) d\textbf{r$^{\prime}$},
\end{equation}
where $U_s^{Approx}(\textbf{r})$  is determined depending on the approximation (Born : $U_s^{Approx}(\textbf{r})$ = $U(\textbf{r}) - U_i(\textbf{r})$ , Rytov : $U_s^{Approx}(\textbf{r})$ = $U_i(\textbf{r}) \log\frac{U(\textbf{r})}{U_i(\textbf{r})}$). Under the assumption of planar incidence, $U_i(\textbf{r}) = e^{i \textbf{k$^{in}$} \cdot \textbf{r}}$, Eq. \ref{eq:Wolf} can be transformed to the following,
\begin{eqnarray}\label{eq:Wolf_FT}
&&\frac{k_z}{2\pi i} \iint_{-\infty}^{+\infty} U_s^{Approx}(\textbf{r};z=0) e^{i(k_xx+k_yy)}dxdy
\nonumber\\
&& = \int_{V} F(\textbf{r$^{\prime}$}) e^{-i(\textbf{k}-\textbf{k$^{in}$}) \cdot \textbf{r$^{\prime}$}} d\textbf{r$^{\prime}$},
\end{eqnarray}
where $U_s^{Approx}(\textbf{r};z=0)$ is the measurement in the image plane and $k_z = \sqrt{k^2 - k_x^2 - k_y^2}$. Eq. (\ref{eq:Wolf_FT}) is equivalent to the 3-D k-space of the object filled with the 2-D Fourier transform of the measured field over the Ewald's sphere.

\subsection{Nonlinear forward model}
We mow describe the BPM based reconstruction algorithm for optical diffraction tomography. We can propagate the light through inhomogeneous medium by splitting the process in multiple fine steps, where each step consists of a diffraction step followed by a refraction step. Denoting the envelope of the wave as $A(\textbf{r})$ for $U(x,y,z) = A(x,y,z)e^{ikz}$, BPM can be written as
\begin{eqnarray}\label{eq:BPM}
A(x,y,z+dz)&& = e^{ik_0\Delta n(x,y,z) dz}
\nonumber\\
 \times F_{2D}^{H}[F_{2D}\{&&A(x,y,z)\} \times e^{-idz\{(k_x^2+k_y^2)/(k+k_z)\}}]
\end{eqnarray}
where $\Delta n$ is the contrast between RI of the sample ($n(x,y,z)$) and $n_0$, $k_z = \sqrt{k^2 - k_x^2 - k_y^2}$. The 2-D Fourier transform operator ($F_{2D}$) is defined as $A(k_x,k_y,z) = \iint A(x,y,z)e^{-i(k_xx+k_yy)}dxdy$ .

\subsection{Iterative reconstruction algorithms}
Once the forward model is determined as either linear or nonlinear, we can specify a cost function that combines an error term and a regularization that incorporates prior knowledge about the sample. We impose the total variation (TV) and the non-negativity constraints \cite{kamilov2016optical}. Specifically, the cost function is defined as
\begin{equation}\label{eq:Cost}
C(f) = \frac{1}{2}||y-A(f)||_2^2 + \tau D(f) + N(f),
\end{equation}
where $y \in \mathbb{C}^{M}$ denotes the experimental measurements, $f \in \mathbb{R}^{N}$ denotes the object function, $A : \mathbb{R}^{N} \rightarrow \mathbb{C}^{M} $ is the forward model which can be either linear or nonlinear, $D(f) = \displaystyle\sum_{n=1}^{N} \sqrt{(\nabla_xf)^2 + (\nabla_yf)^2 + (\nabla_zf)^2}$ ($\nabla_x, \nabla_y, \text{ and } \nabla_z$ are finite difference operators in $x,y \text{ and } z $ dimensions, respectively.), $N(f)$ is the indicator function, $N(f) = \infty$ $(f < 0)$ or $N(f) = 0$ $(f \geq 0)$, and $\tau$ is the regularization parameter, setting the relative weight of the regularization term. To minimize the cost function, Eq. \ref{eq:Cost}. In both cases, we used the Fast Iterative Shrinkage-Thresholding Algorithm (FISTA) \cite{beck2009fast}. For LT, we also rely on the the stochastic gradient method to speed up the computation.

\section{Method}
\subsection{Simulation setup}

\begin{figure}
\centerline{
\includegraphics[height=3.5in]{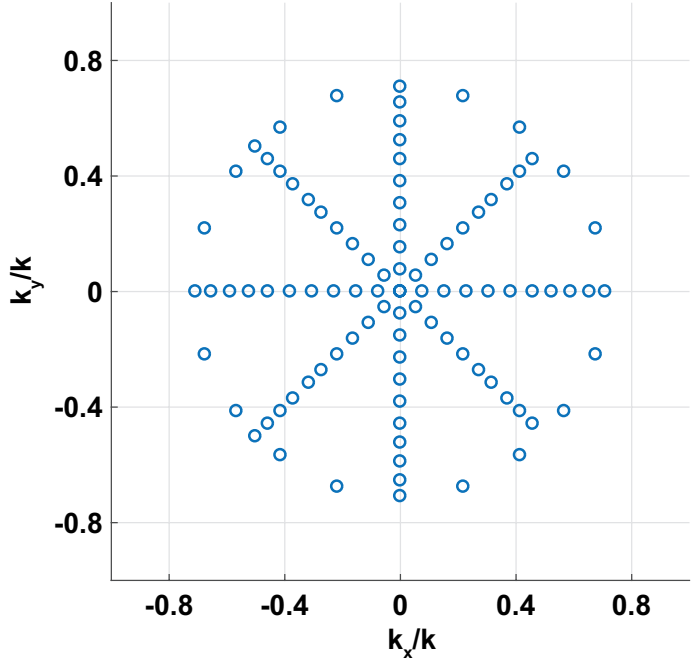}}
\caption{k-space trajectories of normalized illumination k-vectors.}
\label{fig:ksp}
\end{figure}

To obtain the equivalent experimental measurements, we used the Mie theory to derive the scattered field by a single sphere \cite{bohren2008absorption, schafer2012calculation}. The sample was illuminated at 95 different angles and the k-space trajectory of illumination angles is shown in Fig. \ref{fig:ksp}. To investigate the performance of each forward model, we tested the algorithms on following three different schemes. The first is the case of different RI contrasts with fixed radius (Case 1), the second is the case of different radius with fixed RI contrast (Case 2). The third is the case with a fixed sample-induced phase delay but different in both RI contrast and diameter (Case 3). In addition, to investigate the capacity of LT in handling multiple scattering caused by multiple samples, Mie theory for multiple cylinders was used \cite{schafer2012calculation} to investigate both forward models. Uniformly distributed 101 angles between $-\pi/4$ and $\pi/4$ were used to scan the sample. To quantitatively evaluate the performance, we calculated the error defined as,

\begin{equation}\label{eq:RE}
Error(f_{recon}, f_{true}) = \frac{||f_{recon}-f_{true}||_2^2}{||f_{true}||_2^2},
\end{equation}
where $f_{recon}$ is the reconstructed solution and $f_{true}$ is the ground truth from the Mie simulation.
In addition, to compare the relative performances of two forward models, the relative error is defined as,

\begin{equation}\label{eq:RRE}
Relative \text{ } Error(f_{lin},f_{non},f_{true}) = \frac{Error(f_{non}, f_{true})}{Error(f_{lin}, f_{true})},
\end{equation}
where $f_{non}$ and $f_{lin}$ are solutions acquired from the nonlinear and linear models, respectively.

\subsection{Experimental setup}

The experimental apparatus consists of a Mach-Zehnder interferometer. The signal and reference arm are recombined at an angle of 1.43 degrees before the detector in order to record off-axis digital holograms. The light source is a continuous wave laser diode at 406 $nm$ with a coherence length of 250 $\mu m$. The light was spatially filtered and collimated. The sample consisted of two polystyrene microspheres with a nominal diameter of 4.45 $\mu m$. The spheres were placed on two 150 $\mu m$ thick glass cover slips facing each other, each sphere sitting on a different cover slip. The gap between the cover slips was filled with a low fluorescence immersion oil (Nikon type N, n0 = 1.518 at 546.1 $nm$ and ve = 41), with a refractive index of 1.5370 at 406 $nm$. 
The coverslips and spheres assembly was placed in the signal arm between two infinity-corrected 100x oil-immersion microscope objectives (UPlanApo NA1.4 on the detection side and UplanFI NA1.3 on the illumination side). The same oil was used between the objectives and the cover slips. The effective numerical aperture of the system is 1.3. The image of the spheres was projected on a S-CMOS camera (Andor Neo) with an effective magnification of 111. The sphere were illuminated with plane wave through the illumination objective at different incidence angles. This could be achieved by placing two galvo-mirrors (one for the X and the other one for the Y direction) in conjugate image planes of the sample. We acquired 160 views equally spaced on a circle in the $k_{x}-k_{y}$ plane at an angle of 42 degrees from the optical axis (effective angle on the sample).

\subsection{Reconstruction setup}

The algorithms were implemented using custom scripts in MATLAB R2016b (MathWorks Inc., Natick, MA, USA) on a desktop computer (Intel Core i7-6700 CPU, 3.4 GHz, 32 GB RAM). For faster computation, a graphic processing unit (GPU, GeForce GTX 1070) was utilized. The computational space was sampled with the step size ($\Delta x=\Delta y=\Delta z$) of $0.08 \mu m$ (single bead and multiple cylinders) or $0.0856 \mu m$ (two beads). The regularization parameter, $\tau$, was manually set as described in Table 1 ($\Delta n$ : RI contrast, $L$ : diameter in $\mu m$, $\Delta n_0$ : 0.0406, and $l_0$ : 5 $\mu m$). For both models, the step size was reduced after every iteration, $\gamma^{k+1} = 0.985 \gamma^{k}$ (single sphere and two beads) or $\gamma^{k+1} = 0.99 \gamma^{k}$ (multiple cylinders), and the iteration numbers for TV and FISTA are 20 and 200, respectively. In case of LT, the stochastic gradient method was used with 8 randomly chosen angles out of total angles selected at each iteration. For every case, the first order Rytov approximation was used as the initial condition.

\begin{table}[]
\footnotesize
\centering
\caption{Parameters}
\label{my-label}
\begin{tabular}{|c|c|c|c|}
\hline
\begin{tabular}[c]{@{}c@{}}Data (single type)\\ Size (X$\times$Y$\times$Z)\end{tabular}                  & Parameter    & Linear & Nonlinear   \\ \hline
\multirow{2}{*}{\begin{tabular}[c]{@{}c@{}}Single sphere\\ (350$\times$350$\times$128)\end{tabular}}     & $\gamma$     & 0.01   & 0.001        \\ \cline{2-4} 
                                                                                           & $\tau$       & \text{$\tau = 2 \Delta n \times L / (\Delta n_0 \times l_0)$}      & 0.1 \\ \hline
\multirow{2}{*}{\begin{tabular}[c]{@{}c@{}}Multiple cylinders\\ (1024$\times$256)\end{tabular}}   & $\gamma$     & 0.01   & 0.001 \\ \cline{2-4} 
                                                                                           & $\tau$       & 0.4   & 0.1 \\ \hline
																																													\multirow{2}{*}{\begin{tabular}[c]{@{}c@{}}Two beads \\ (256$\times$256$\times$256)\end{tabular}}   & $\gamma$     & 0.01   & 0.001 \\ \cline{2-4} 
                                                                                           & $\tau$       & 4   & 1.5 \\ \hline
\end{tabular}
\end{table}

\section{Results}

\subsection{Case 1}
\begin{figure}
\centerline{
\includegraphics[width=4in]{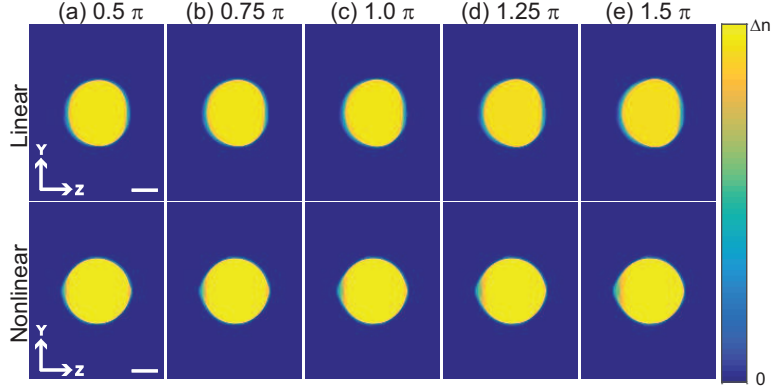}}
\caption{YZ cross sections of RI tomograms of single bead using two different forward models. Upper row: linear (Rytov), lower row: nonlinear (BPM). From (a) to (e), the RI contrast proportionally increase so that the sample-induced phase delay equals to the number written on each column. Scale bars are 2 $ \mu m $.}
\label{fig:IMG_dn}
\end{figure}

\begin{figure}
\centerline{
\includegraphics[width=4in]{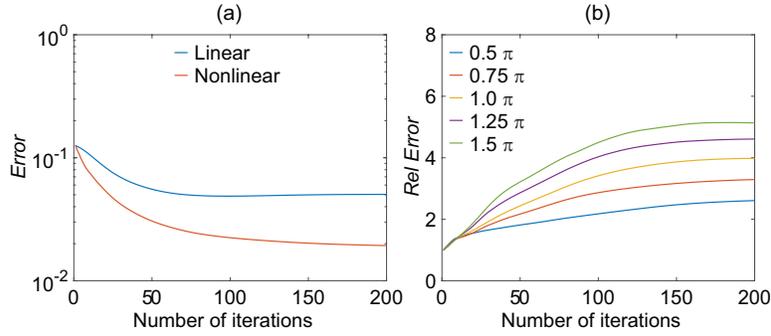}}
\caption{(a) $Error$ of the linear and nonlinear models for 5 $ \mu m$ bead inducing 0.5 $\pi$ phase delay. (b) $Relative \text{ } Error$s for various beads which differ in the RI contrast.}
\label{fig:RRE_dn}
\end{figure}

Case 1 shows the effect of the RI contrast on the performance of the linear and nonlinear models. While both models show gradual distortions with the increase of the RI contrast in each case, the nonlinear model outperforms the linear one (Fig. \ref{fig:IMG_dn}). This is quantified by the $Error$ curve in Fig. \ref{fig:RRE_dn}(a), showing lower $Error$ of the nonlinear model for all cases (For the other cases, $Error$s are in the appendix). The relative effect of the increase in RI contrast can be investigated by looking at $Relative \text{ } Error$ (Fig. \ref{fig:RRE_dn}(b)). As the RI contrast increases, $Relative \text{ } Error$ increases in the range from 0.5 $\pi$ to 1.5 $\pi$. 

\subsection{Case 2}
\begin{figure}
\centerline{
\includegraphics[width=4in]{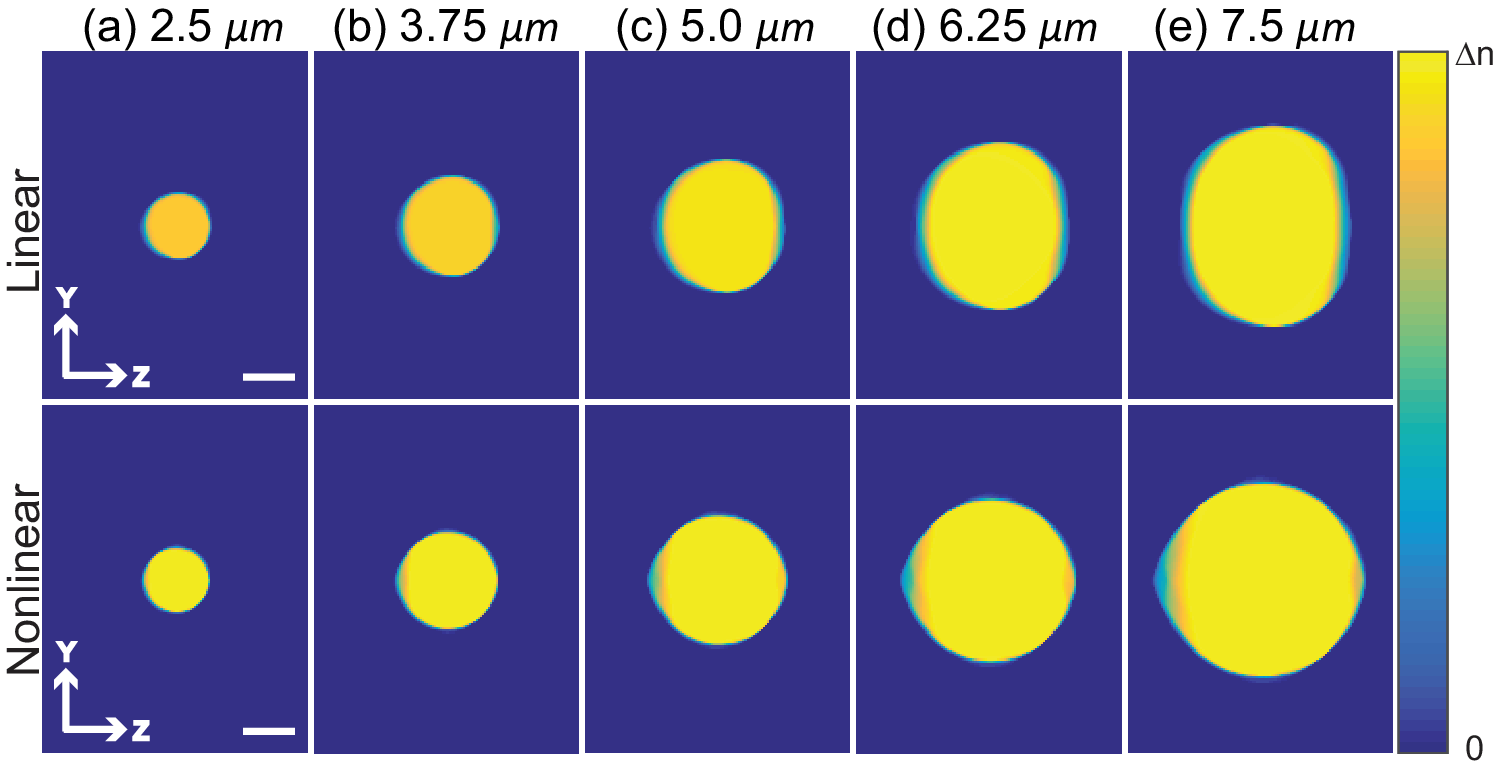}}
\caption{YZ cross sections of reconstructed RI tomograms of single bead using two different forward models. Upper row: linear (Rytov), lower row: nonlinear (BPM). From (a) to (e), the diameter proportionally increases. Each diameter is written on each column. Scale bars are 2 $ \mu m $ and $\Delta n$ is 0.0406.}
\label{fig:IMG_dl}
\end{figure}

For Case 2, we assessed the validity of the models by increasing the diameter of sphere. It can be clearly observed that the reconstruction results from the linear model tend to shrink along the optical axis with the increase of the diameter (Fig. \ref{fig:IMG_dl}). In contrast, the nonlinear model shows consistent tomograms (Fig. \ref{fig:IMG_dl}) and lower $Error$ values than that of the linear model (Fig. \ref{fig:RRE_dl}(a), for the other cases, $Error$s are in the supplementary materials). The advantage of LT against linear tomography can be clearly seen in Fig. \ref{fig:RRE_dl}(b). We can observe the gradual increase of $Relative \text{ } Error$ values with the gradual increase of the diameter, indicating the gradual break-down of the linear model.
To be specific, by looking at the converging points of each $Relative \text{ } Error$, it can be observed that the incremental change in $Relative \text{ } Error$ gets bigger with the increase of the diameter. 

\begin{figure}
\centerline{
\includegraphics[width=4in]{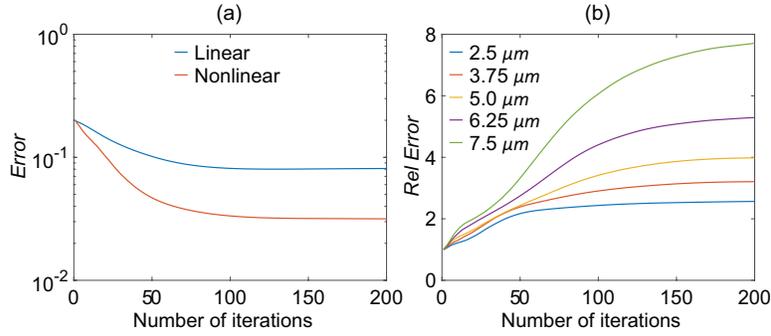}}
\caption{(a) $Error$ of the linear and nonlinear models for 2.5 $ \mu m$ bead inducing 0.5 $\pi$ phase delay. (b) $Relative \text{ } Error$s for various beads which differ in the diameter.}
\label{fig:RRE_dl}
\end{figure}

\subsection{Case 3}
The purpose of Case 3 is to see the relative importance of two different factors, the RI contrast and the size of the sample. Here, the size of sphere varies keeping the sample-induced phase delay to $\pi$. Therefore, each sample differs in both RI contrast and size, but the product of the two factors remains the same. It is obvious that LT outperforms linear tomography as shown in Fig. \ref{fig:IMG_dldn} and Fig. \ref{fig:RRE_dldn}(a) (For the other cases, $Error$s are in the appendix). In Case 3 where the two factors have compensatory effects on $Error$s, it was possible to observe that $Relative \text{ } Error$ values for the cases of 2.5 $\mu m$, 3.75 $\mu m$ and 5 $\mu m$ converge (Fig. \ref{fig:RRE_dldn}(b)) with the number of iteration.

\begin{figure}
\centerline{
\includegraphics[width=4in]{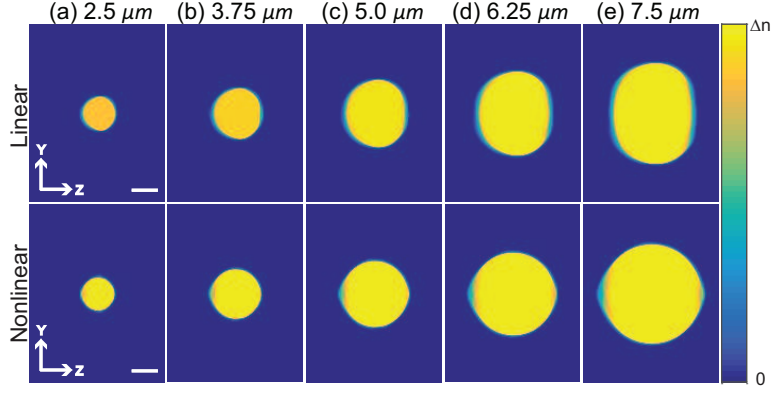}}
\caption{YZ cross sections of RI tomograms of single bead using two different forward models. Upper row: linear (Rytov), lower row: nonlinear (BPM). From (a) to (e), the diameter proportionally increases, the RI contrast decreases keeping sample-induced phase delay $\pi$. Each diameter is written on each column. Scale bars are 2 $ \mu m $.}
\label{fig:IMG_dldn}
\end{figure}

\begin{figure}
\centerline{
\includegraphics[width=4in]{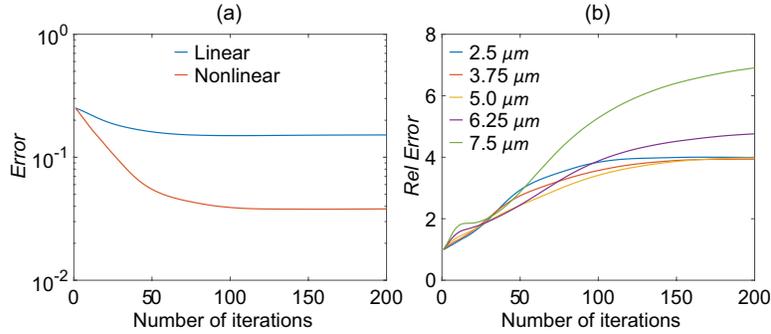}}
\caption{(a) $Error$ of the linear and nonlinear models for 2.5 $ \mu m$ bead inducing 1 $\pi$ phase delay. (b) $Relative \text{ } Error$s for various beads which differ in both the RI contrast and the diameter.}
\label{fig:RRE_dldn}
\end{figure}

\subsection{Comparative analysis}
\begin{figure}[h]
\centerline{
\includegraphics[width=4in]{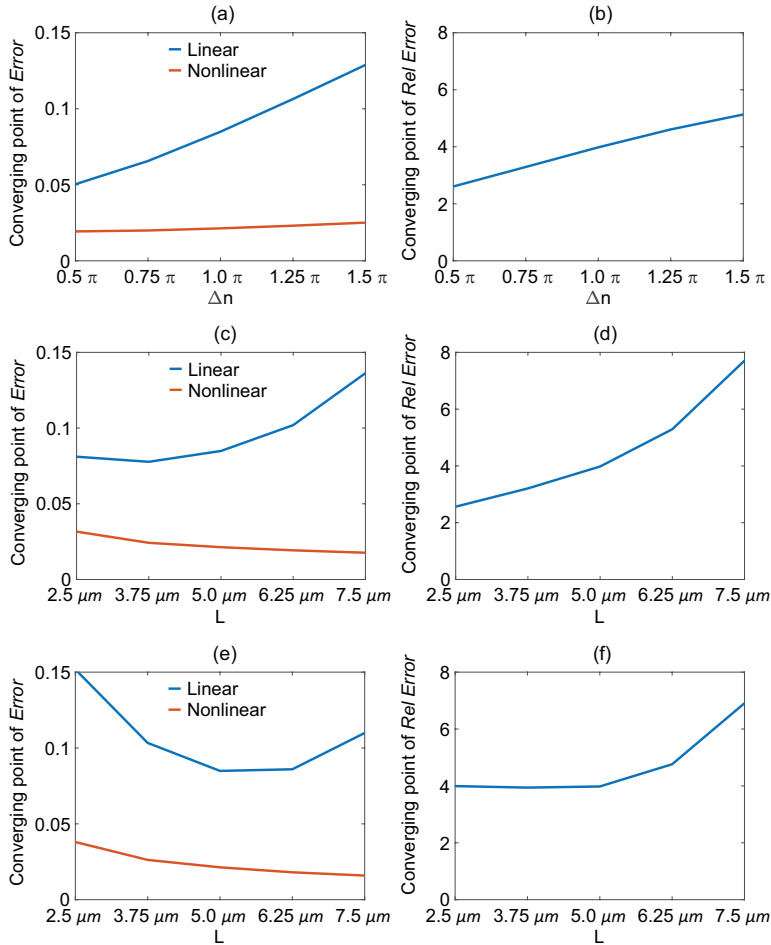}}
\caption{Converging points of $Error$s and $Relative \text{ } Error$s for Case 1 ((a) and (b)), Case 2 ((c) and (d)) and Case 3 ((e) and (f)).}
\label{fig:Converging}
\end{figure}

The converging points of $Error$s and $Relative \text{ } Error$s are plotted in Fig. \ref{fig:Converging}. Fig. \ref{fig:Converging}(a,c,e) not only show overall tendencies but also give us more information about how $Error$s converge depending on each individual factor explaining changes in $Relative \text{ } Error$s as shown in Fig. \ref{fig:Converging}(b,d,f). Overall, linear model breaks down with either the increase of the RI contrast or the size, but the dependence on each individual factor shows different patterns as you can see from the blue curves in Fig. \ref{fig:Converging}(a,c). Compared to that, LT shows much less changes in converging points (red curves in Fig. \ref{fig:Converging}(a,c)). While it shows positive slope in Fig. \ref{fig:Converging}(a), interestingly, the slope remains negative even when the size increases. These effects result in different dependencies of $Relative \text{ } Error$s as in \ref{fig:Converging}(b,d). When both of the RI contrast and the size change (Fig. \ref{fig:Converging}(e,f), it become more complex. Converging points of $Relative \text{ } Error$s almost stay constant in the range from 0.5 $l_0$ to 1 $l_0$, and increase fast after 1 $l_0$ as shown in Fig. \ref{fig:Converging}(f). It comes from the different effectiveness of each factor as shown in Fig. \ref{fig:Converging}(e). To be specific, in terms of $Relative \text{ } Error$s, when the RI contrast increases and the size decreases, the effect of each factor compensates the opposite effect of the other factor relatively showing small changes in the converging points of $Relative \text{ } Error$. However, the effect of the increase in size is stronger than the effect of the decrease in the RI contrast with the increase in the size over the size, $l_0$.

\subsection{Multiple cylinders}
\begin{figure}[h]
\centerline{
\includegraphics[width=4in]{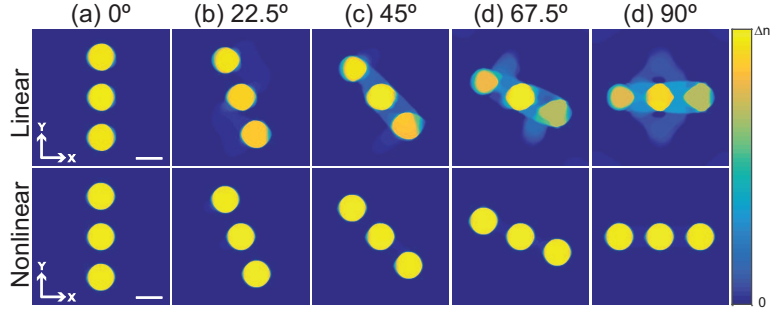}}
\caption{YX cross sections of RI tomograms of multiple cylinders using two different forward models. Upper row : linear, lower row : nonlinear. From (a) to (e), the sample is gradually rotated from $0$ degree to $90$ degrees. Scale bars are 4 $ \mu m $ and $\Delta n$ is 0.0508.}
\label{fig:MC}
\end{figure}

\begin{figure}[h]
\centerline{
\includegraphics[width=4in]{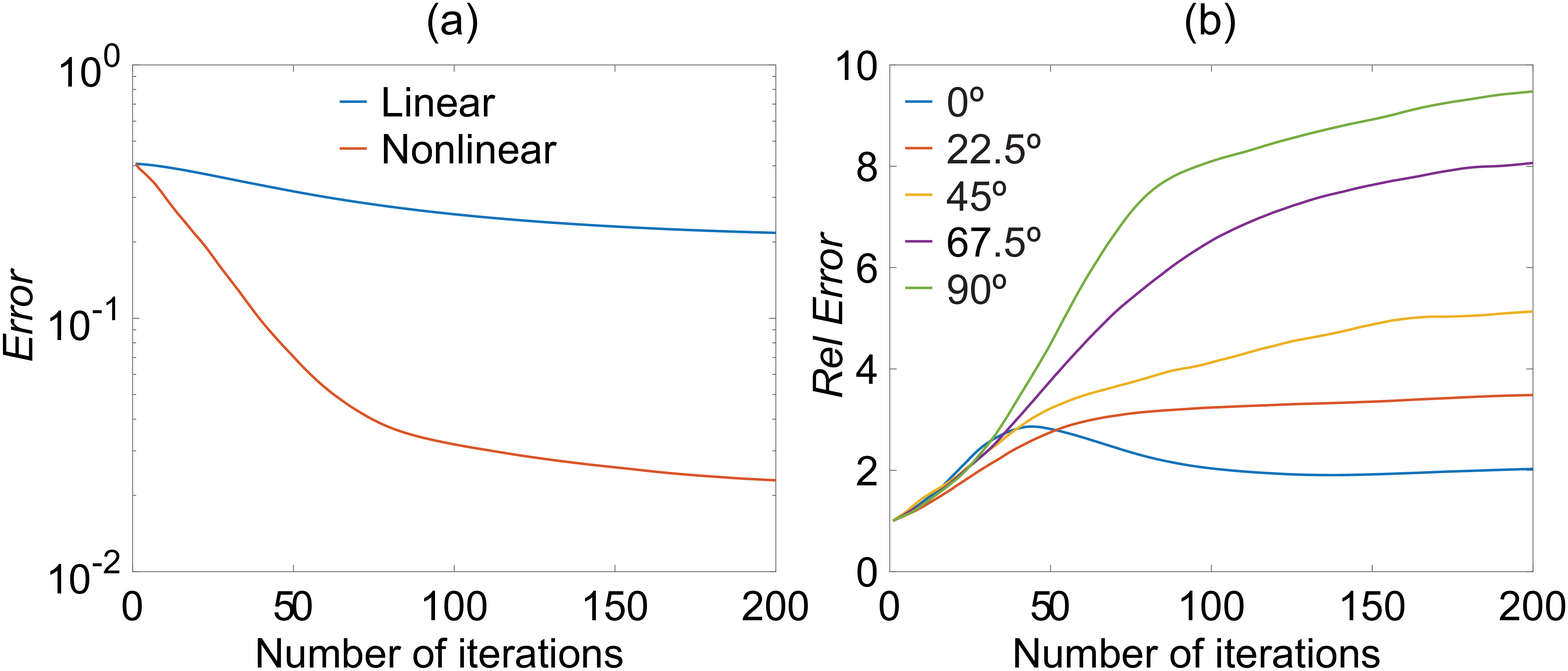}}
\caption{(a) $Error$ of the linear and nonlinear models for multiple cylinders rotated 0 degree. (b) $Relative \text{ } Error$s for multiple cylinders rotated from 0 degree to 90 degrees.}
\label{fig:MC_converging}
\end{figure}

Even though each cylinder is a week scatterer, when they are aggregated, multiple scattering caused among them can introduce severe distortions \cite{azimi1983distortion}. To test the capacity for handling multiple scattering, simulated measurements for multiple cylinders has been generated using the Mie theory \cite{schafer2012calculation}. A set of three cylinders whose diameter is 4 $ \mu m $ and RI value is $ 0.0508 $ were considered. Then, the set of cylinders was rotated from 0 degree to 90 degrees resulting in the gradual increase of the multiple scattering effect. As shown in Fig. \ref{fig:MC}, we can clearly see that LT outperforms the linear model for all cases. While the linear tomography shows leakage between cylinders, the LT maintains a clear distinction of three cylinders. It can be quantitatively confirmed through the converging points of $Error$s and $Relative \text{ } Error$s, as shown in Fig. \ref{fig:MC_converging}.

\subsection{Phase unwrapping}
\begin{figure}[h]
\centerline{
\includegraphics[width=4in]{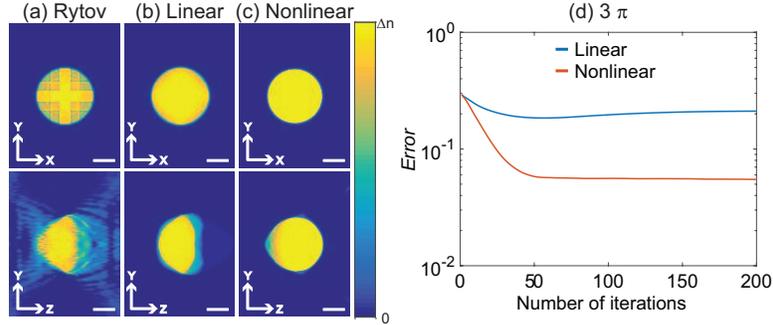}}
\caption{(a) : XY and YZ cross sections of RI tomograms of the high contrast bead from Rytov approximation. (b-c) : Two forwards model were used to reconstruct RI tomograms, (b) linear tomography and (c) LT. (d) $Error$ of linear and nonlinear models. Scale bars are 2 $ \mu m $ and $\Delta n$ is 0.1218.}
\label{fig:Unwrap}
\end{figure}

The Rytov algorithm makes use of calculated on the measured phase of each projection. If the optical thickness of the sample exceeds 2 $\pi$ then the measured phase must be unwrapped. Here, we compare the performance of these models when the phase unwrapping algorithm fails. The RI contrast is increased so that sample-induced phase delay resulted in 3 $\pi$. It is then difficult to properly unwrap the phase, even with a state-of-the-art algorithm such as a PUMA which was used throughout this paper \cite{bioucas2007phase}. Since the Rytov approximation requires unwrapped phases for the reconstruction, the failure in unwrapping directly relates to severe distortion in the reconstruction result of Rytov approximation, as shown in Fig. \ref{fig:Unwrap}(a). It is critical not only to linear tomography, which uses the linear model based on Rytov approximation, but also to the LT because it is more likely for LT to fall in a local minimum. Compared to the linear tomography which only shows denoising effects (Fig. \ref{fig:Unwrap}(b)), the LT is able to successfully reconstruct the tomograms (Fig. \ref{fig:Unwrap}(c)) even though it uses \ref{fig:Unwrap}(a) as the initialization. The effect is more dramatic in terms of $Error$ as in Fig. \ref{fig:Unwrap}(d).

\subsection{Experimental results}

\begin{figure}[h]
\centerline{
\includegraphics[width=4in]{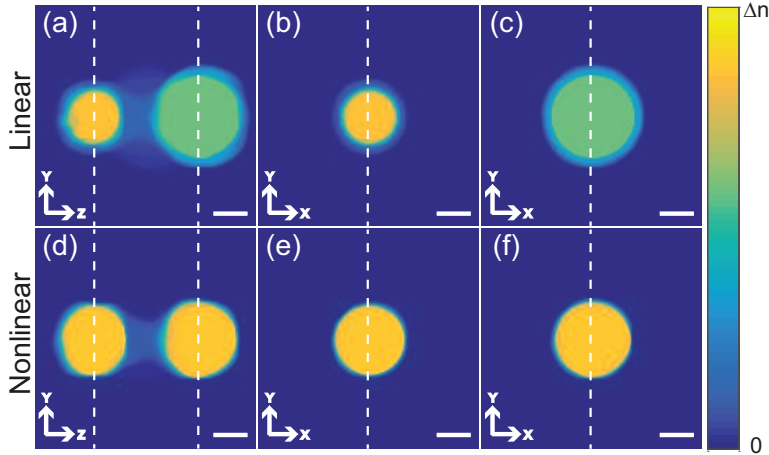}}
\caption{
 YZ ((a) and (d)) and XY ((b) and (e) for the left bead, (c) and (f) for the right bead) cross sections of RI tomograms of two beads using two different forward models. Upper row: linear (Rytov), lower row: nonlinear (BPM). The white dotted lines represent the slices of the complimentary figures. Scale bars are 2 µm and $\Delta n$ is 0.0972.}
\label{fig:Twobeads}
\end{figure}

We conducted an experiment to assess LT. Two 4.45 $ \mu m $ were placed in a row so they overlapped in z-axis. We applied Linear tomography and LT on an experimental data. In case of linear tomography which is based on Rytov, a location of the image plane is very important. Fig. \ref{fig:MC}(c) confirms this fact. As the sample is placed farther from the image plane, RI tomograms gets either smaller in size and higher in RI contrast or bigger in size and lower in RI contrast. It was experimentally confirmed through 2 beads. It comes from the fact that Rytov directly uses phase perturbation in the field. As shown in Fig. \ref{fig:Twobeads}(a-c), one below the image plane gets smaller in size and higher in RI contrast and the other above the image plane gets bigger in size and lower in RI contrast. By contrast, LT clearly reconstructs two beads of equal size and contrast (Fig. \ref{fig:Twobeads}(d-f)). Since BPM describes the propagated field itself through the sample, it is not only unaffected by the location of the image plane but also able to handle multiple scattering.

\section{Conclusion}
In this paper, we rigorously compare the LT against the conventional linear tomography. Mie theory provided the analytical solution for the scattered filed given a sphere so that we were able to evaluate the reconstruction fidelity of each model accurately. To investigate the capability of each model in dealing with nonlinearity, two factors which are directly related to nonlinearity, the RI contrast and the size, were controlled either independently or simultaneously. In all the cases, LT consistently outperformed linear reconstruction methods. We attribute this to the fact that the BPM used by LT captures multiple forward scattering events. The improvement in performance becomes more pronounced as the index or the diameter of the beads increases.  Our results indicate that LT provides a more pronounced improvement in the image quality for large size beads. The most dramatic improvement was observed when we imaged multiple objects (three cylinders). This confirms the observation that object size matters since we can consider  the set of 3 cylinders set as a single large object. In general the BPM performs relatively  well in the simulation of inhomogeneous media with small index contrast since the main limitation of the method is the assumption that reflections can be neglected. For samples whose optical path exceeds 2 $\pi$, phase unwrapping must be deployed in the Rytov algorithm. When the phase unwrapping algorithm fails as the optical path across the 5 $\mu m$ sphere increased to 3 $ \pi$, the linear reconstruction becomes severely distorted. For this case, we observed that the iterations of the LT algorithm were able to correct the distortions that are evident in the Rytov reconstruction due to the phase unwrapping limitations. Finally, the reconstruction of the Rytov algorithm is in focus only at the plane of best focus of the optical system [?,our future paper]. For thick samples the sample becomes blurred away from the focal plane. This is evident in Figures \ref{fig:MC}(a-e) and \ref{fig:Twobeads}(a-c). The distortion evident in Fig. \ref{fig:MC}(c) for example is a combination of two factors: Defocussing and multiple scattering. The BPM helps alleviate both of these problems since it takes multiple scattering into account  and it allows us to keep the entire sample in focus. It was experimentally confirmed using 2 Beads overlapping in the optical axis. As the sample becomes more complex (thicker, higher index contrast) ultimately the BPM provides an inadequate estimate for the scattered field by the object since reflections were neglected and the vectorial nature of the optical field was ignored. In this case the only way to realize an improvement in performance is by adopting a more sophisticated scattering model. There is an intermediate level of sample complexity however where BPM still provides a reasonably accurate prediction of the scattered field but the nonlinear inversion problem becomes very difficult due to emergence of strong local minima. In this regime we believe that there is a global minimum which is a good approximation of the true object but the LT algorithm cannot find it. It is possible that more powerful optimization algorithms than the stochastic decent we used in this paper can provide significant improvement in performance.

\newpage

\appendix
\section{Supplementary figures}
As supplementary figures, $Error$s for each sample for each case are provided in detail.

\begin{figure*}[h]
\includegraphics[width=5 in]{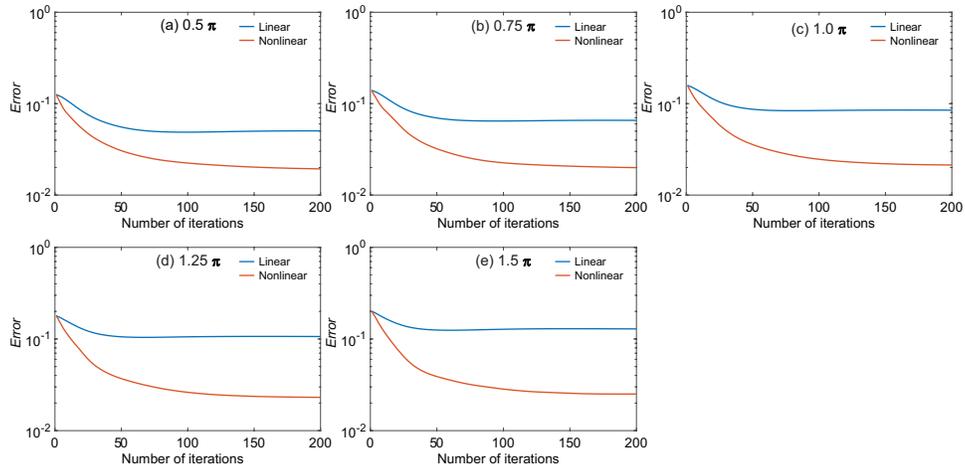}
\caption{$Error$s of linear and nonlinear models for various beads which differ in the RI contrast. From (a) to (f), the RI contrast proportionally increases so that sample-induced phase delay equals to the number written on each figure.}
\label{fig:RE_dn}
\end{figure*}

\begin{figure*}[h]
\includegraphics[width=5 in]{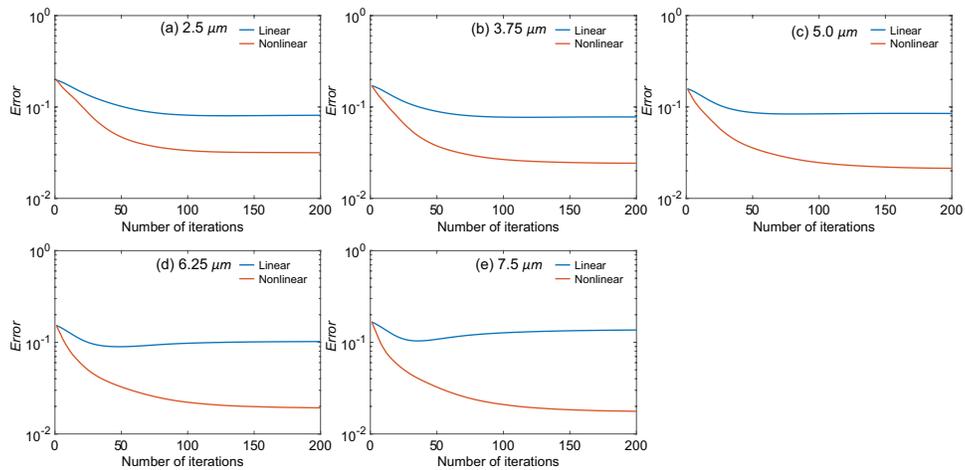}
\caption{$Error$s of linear and nonlinear models for various beads which differ in the diameter. From (a) to (e), the diameter proportionally increases. Each diameter is written on each figure.}
\label{fig:RE_dl}
\end{figure*}

\begin{figure*}[h]
\includegraphics[width=5 in]{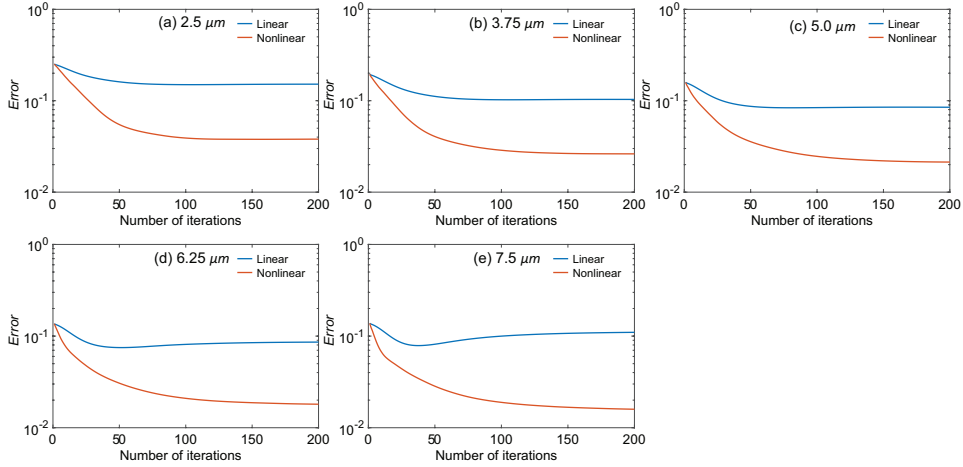}
\caption{$Error$s of linear and nonlinear models for various beads which differ in the diameter. From (a) to (e), the diameter proportionally increases, the RI contrast decreases keeping sample-induced phase delay 1 $\pi$. Each diameter is written on each figure.
}
\label{fig:RE_dldn}
\end{figure*}

\begin{figure*}[h]
\includegraphics[width=5 in]{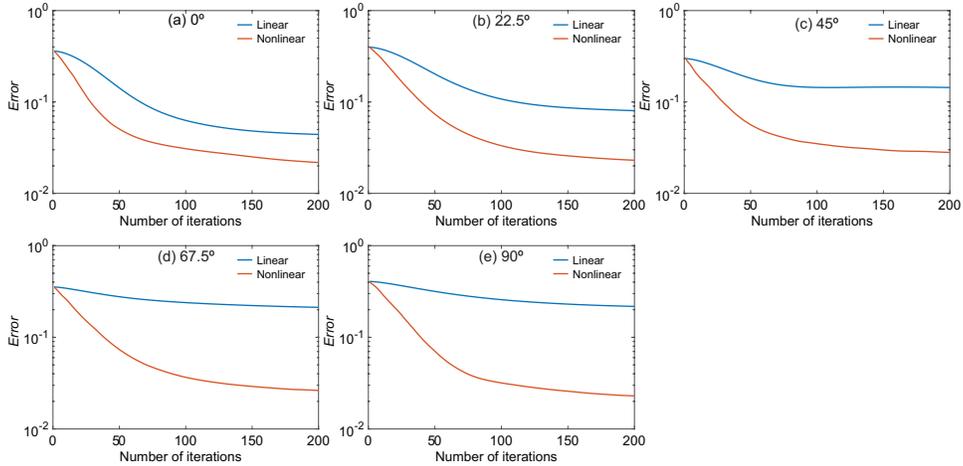}
\caption{$Error$s of linear and nonlinear models for cylinders which rotate from 0 degree (a) to 90 degrees (e) with respect to the optical axis.
}
\label{fig:RE_dAng}
\end{figure*}

\newpage

\bibliography{my_bib_abbr}

\end{document}